\documentclass[10pt, titlepage, twocolumn]{report}

\usepackage{titlesec}
\usepackage{fancyhdr}
\usepackage{lipsum}
\usepackage{tikz}
\usepackage{pgfplots}
\usepackage{siunitx}
\usepackage{pgfplotstable}
\usepackage{graphicx}
\usepackage{mathtools}
\usepackage[english]{babel}
\usepackage[printwatermark]{xwatermark}
\usepackage{xcolor}

\usepackage[hidelinks]{hyperref}

\usepackage[labelfont=bf]{caption}
\usepackage[T1]{fontenc}
\usepackage{blindtext}
\usepackage{amssymb}
\newcommand*{\field}[1]{\mathbb{#1}}%

\usepackage{amsmath}
\usepackage{algpseudocode,algorithm,algorithmicx}
\usepackage{fancyvrb}
\usepackage{geometry}
\usepackage[utf8]{inputenc}

\graphicspath{ {/} }

\titleformat{\section}
{\normalfont\scshape\centering}{\thesection}{1em}{}

\titleformat{\section}
{\normalfont\Large\bfseries}{\thesection}{0em}{}[{\titlerule[0.8pt]}]

\titlespacing\section{0pt}{12pt plus 4pt minus 2pt}{10pt plus 2pt minus 2pt}

\date{\today}

\date{}

\titleformat{\section}
  {\normalfont\scshape\centering}{\thesection}{0em}{}
  
\titleformat{\section}
  {\normalfont\Large\bfseries}{\thesection}{1em}{}[{\titlerule[0.8pt]}]

\makeatletter
\renewcommand\thesection{}
\renewcommand\thesubsection{\@arabic\c@section.\@arabic\c@subsection}
\makeatother

\begin{document}
\begin{titlepage}

\protect\parbox{\textwidth}
	{\protect\centering 
		\huge AfricaOS:  Using a distributed, proposal-based, \textit{replicated state machine} towards \textit{liberation} from the Berlin Conference of 1885 
	} 
\begin{center}
{\LARGE
\today}
\end{center}

\vfill
{
\Large
\begin{center}
\texttt{Jovonni L. Pharr} \\
\texttt{jpharr2@student.gsu.edu} \\
\texttt{Atlanta, GA, USA} \\
\texttt{[v0.0.1]} \\
\end{center}
}

\setlength\parindent{24pt}

\vfill
\section*{Abstract}
\noindent

The \textit{Berlin Conference} of 1885 has influenced the way native Africans, and the African Diaspora live their daily lives. France contractually controls several resources generated by the continent of Africa. Herein lies a technical proposal to free Africa from the financial and economic agreements coerced upon the continent over a century ago by utilizing decentralized collaboration through advanced technology. AfricaOS (\(\texttt{AOS}\)) aims to provide a philosophical, and fundamental framework for implementing a simple, distributed, collaborative computer for agreement amongst peers. The work also demonstrates an algebra over transactions, use of the protocol for privatization, a method for tokenizing \textit{barter economies}, and methods to design mechanisms for use in implementing protocol behavior. 

\end{titlepage}

\tableofcontents
\listoffigures

\clearpage

\onecolumn
\section{Foreword}
\hspace*{24pt}
The effects of the \textit{Berlin Conference of 1885} on the continent of Africa, and its diaspora have been studied by scholars for decades. Colonization of any environment impacts the future of that area, its people, and the posterity of its people. Financial impact of the Berlin Conference has led to the manipulation of the financial future of the entire continent. A large group of West African countries plan to adopt a single fiat currency, called the \textit{Eco}, thus reforming the CFA Franc. 

\hspace*{24pt}
The Eco is an attempt to free the continent from effects of colonialism as Paris will no longer co-manage the currency. Foreign exchange reserves will no longer be centralized by France and the obligation to pay 50\% of these reserves to the French Treasury operating account will no longer be in effect. However, with the current geopolitical climate, and zeitgeist, implementing a fiat currency that is centralized comes with a set of problems of which can have the same level of impact as colonialism. The \textit{Banque de France} will remain the guarantor and regulator of converting the Eco to the Euro. They will also maintain a fixed parity between the two currencies. This centralized party will still have a large influence on the long-term ramifications of this deal. 

\hspace*{24pt}
With the advancements today, the entire continent of Africa can use modern financial technologies to circumvent these centralized concerns. It is my belief that it makes more practical sense, is more sustainable in the long-term, and future-proof to implement an open, neutral, decentralized, borderless, and censorship resistant digital currencies to sever the long-standing, economic grasp of colonialism on the continent of Africa. Herein lies a proposed, distributed, operating system with which to power the next generation of African economic empowerment. This system can be used toward liberation of any economy under the same economic grasp, but the continent is a perfect initial testing bed for this focused mission.

\hspace*{24pt}
At the time of writing, the \textit{Novel Corona Virus}, COVID-19, has not only been spreading across the planet, but has also been bringing unprecedented challenges to several countries. The economic state of the world is under an attack by a novel, invisible enemy. Governments are putting forth stimulus packages, and attempting to inject money into their respective systems, and deliver money directly to their constituents. The price of Bitcoin has been nearly cut in half since the number of corona virus cases has spread so widely. It is at times like this where we realize the amount of reliance the general public has on government programs, and the currency of their nation. In my opinion, this will mark a changing point in our economic expectations of which will be similar to the time Satoshi Nakamoto introduced the Bitcoin project. Populations need to regain sovereignty from their nation's money, and must also obtain the capability to introduce their own supported means of economic transactions into the world without the need to spend years on research \& development.

\begin{quotation}
"For to be free is not merely to cast off one's chains, but to live in a way that respects and enhances the \textit{freedom} of others." - \texttt{Nelson Mandela}
\end{quotation}

\begin{figure}[ht]
\begin{flushright}
	\includegraphics[width=0.2\textwidth]{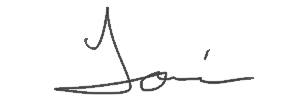}	
	\end{flushright}
\end{figure}

\clearpage
\twocolumn

\section{Background}

\hspace*{15pt}
This protocol is the first direct attempt to practically provide a technological framework for alleviating the financial ramifications brought about by the Berlin Conference of 1885. 

\subsection{The Berlin Conference}
\hspace*{15pt}
Otherwise known as the West African Conference, the Berlin Conference \cite{A1}\cite{A2}\cite{A3} is among the most important events in African History influencing the lives of many today. Historical scholars have attributed much of the formalized colonization of Africa to this event. This was not the only attempt at formalizing Africa's colonization, but has contributed to colonization becoming less ad hoc, and more organized among colonialists.

\section{Liquidity}
\hspace*{15pt}
There exists a liquidity problem in several "impoverished" countries on the continent. I use the term "impoverished" because although the environment may not facilitate a large amount of fiat-based economic trade, it may sit upon a wealth of natural resources. With Africa being one of the most resource-rich places on the planet, there should not exist such a large and widespread problem of liquidity. 

\subsection{The Problem}
\hspace*{15pt}
The problem exists where many individuals simply do not have access to money. Additionally, if money does exists, it may be under the control of externalities. With the advent of digital currencies, lack of implementation is the only reason why an independent environment cannot create a currency of their own to facilitate economic activity. With the internet, the environment can even expose access to the created currency to the outside world, and decide upon the distributed rules governing the environment's economy. 

\hspace*{15pt}
With colonial influence on the economy of several countries, France possesses the power to manipulate the french-based fiat currencies used by french-colonized countries. Today, France is granted financial reserves from African countries of which is used within the french financial markets. These reserves given to France are not allowed to be retrieved in full by the African countries, but are only allowed to be borrowed against as a loan, at a specified interest rate.

\begin{figure}[ht]
\centering
	\includegraphics[width=0.5\textwidth]{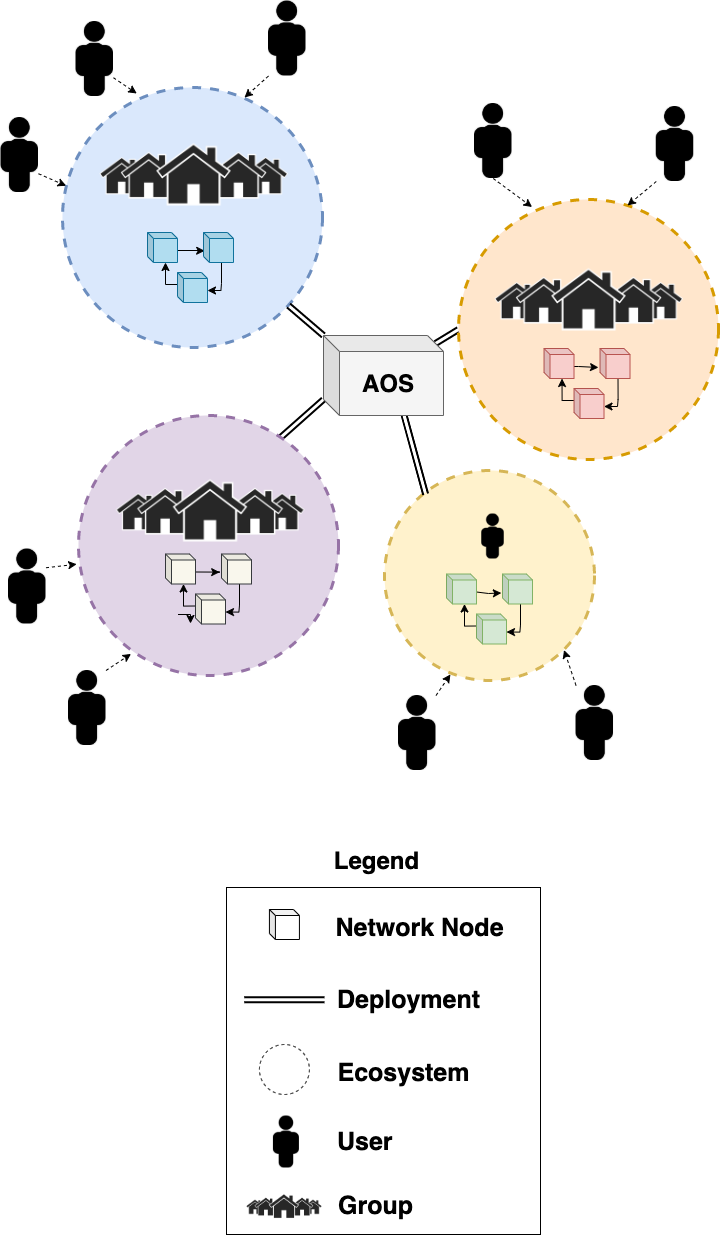}
	\caption{An (\texttt{AOS}) setup, exposing multiple separate instances of ecosystems, with separate networks}
	\label{aos}
\end{figure}

\section{Market Creation}
\hspace*{15pt}
AfricaOS (\(\texttt{AOS}\)), illustrated in figure \ref{aos}, extends the ability for any one person, group of people, or entities, to foster their own markets with their own market rules. Bitcoin enabled a person, or group to create their own currency, as long as they could provide or garner nodes to support their network. However, the underlying logic for bitcoin is difficult to change, and not friendly to non-technologists. Over time, developers have figured out ways to alter Bitcoin to meet their needs, and even going as far as changing core functionality. \(\texttt{AOS}\) contains mechanisms to go beyond the generation of a currency by abstracting away the purpose of the collaboration, and how value is derived in the system.

\hspace*{15pt}
A valuable currency underlying distributed collaboration provides incentive for participants to elicit desired behavior within the system. Underpinning a network with a currency has its benefits, although economic transactions may not be the sole purpose of the network.

\hspace*{15pt}
An effect of market creation is enabling individuals to create markets within which they themselves can participate, the goal being income generation, and facilitating trade. Moreover, market creation, and participation allows for someone to move their income from the labor market, to the capital market.

\section{Mobile Money}
\hspace*{15pt}
Mobile Money \cite{MM} is the concept of individuals having access to banking services through simple, non internet-enabled phones. In more developed ecosystems, a person may have access to a smart phone, and a mobile application for a bank. 

\hspace*{15pt}
Mobile money allows any mobile subscriber to add credit to his or her mobile account and store it for later use or send it to other mobile subscribers via \textit{short messaging service} SMS. The receiver can inexpensively convert this credit back into cash. Mobile money allows users to send cash as quickly as a text message, avoiding inconvenient and costly transfer methods such as physical travel, the mail, or traditional money transfer services like Western Union \cite{MM}.

\hspace*{15pt}
Through mobile money, a person can facilitate their banking needs on-the-go. For less developed environments, where smart phones are not abundant, mobile money allows for an equal amount of banking opportunity. Several countries in Africa have already adopted several uses for mobile money in practice.

\subsection{Mobile Minutes}
\hspace*{15pt}
Mobile minutes are purchased by individuals to support simple telecommunications activities, such as calling another phone number. However, telecommunications companies have began to purchase remaining mobile minutes from \textit{minute holders}, for the purpose of "reselling" them back to the public. This has inadvertently created market demand in which individuals facilitate trade amongst each other since possessing mobile minutes yields economic opportunity.

\subsubsection{Economic Impact}
\hspace*{15pt}
Today, individual vendors, and sellers of goods accept mobile minutes as payment, because they are liquid. This form of "money" contributes to these environments being perfect testing beds for innovation regarding digital currencies. Today, to the average user of mobile money in these countries, using a digital currency can seem natural.

\subsection{SMS}
\hspace*{15pt}
Mobile money, is still a widely used medium of commerce on the continent. To enable easier commerce, nodes within the network are supplemented with \textit{SIP trunks} to accept SMS messages as transaction invocations. 

\hspace*{15pt}
For example, if \(\texttt{Alice}\) wants to send a transaction to \(\texttt{Bob}\), she can send an SMS message to the network, from which a transaction is constructed. There is a simple mapping of addresses to numbers, upon which more features are to be added. This process also brings about the need for secure, and reliable address-phone number mapping methods. The entry point for \textit{SMS} is not a core protocol feature, but directly addresses internet requirements for the proposed system to function, and remain "available" to network participants.

\section{Tokenized Bartering}

\begin{figure*}[ht]
\centering
	\includegraphics[width=\textwidth]{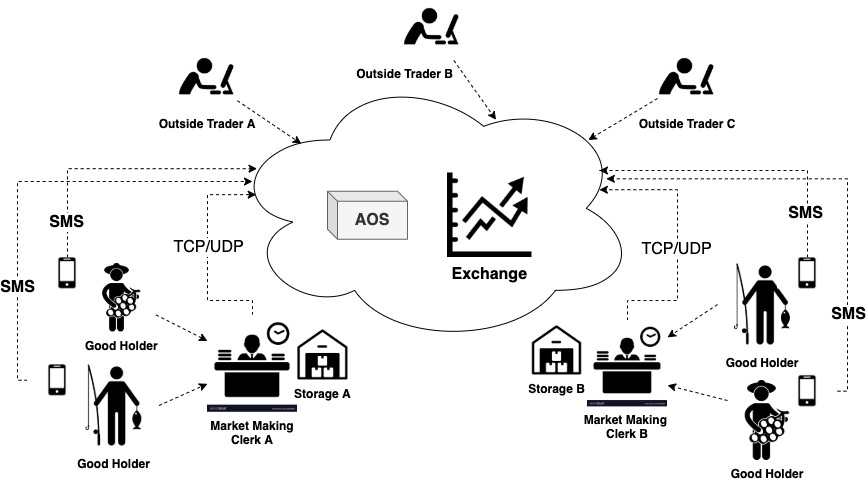}
	\caption{Entry workflow for good holders to tokenize goods for digital, economic exchange}
	\label{bartering}
\end{figure*}

\hspace*{15pt}
While bartering, sellers perform economic transactions within a physical marketplace. However, goods can only be sold to other buyers in the same physical marketplace. Depending on the type of good, surpluses may result in perished items from the sellers perspective. These perished items may result in guaranteed financial loss for the seller.

\subsection{Tokenizing a good}
\hspace*{15pt} 
Shown in figure \ref{bartering}, a member owning a set of coins can exchange them with anyone, for any good they wish to obtain. If an individual has a good, but no coins, they will not be able to participate in the underlying market. Gaining entry into the market requires a member to obtain units of an asset represented within the market. A member with no coins, can exchange their goods with a coin-possessing member, and thus gain entry. This is similar to someone exchanging a digital currency for anything they see as valuable. In this sense the coin holder is shown as a market making clerk. The method through which a market making clerk obtains coins is arbitrary. They can earn coins, or have coins given to them by coin holders.

\hspace*{15pt}
By providing exchange for a market's currency, the member who sells their coins becomes a gatekeeper to the market, for a \textit{coinless}, good holder. This enables market participants to setup businesses purposed with providing access to market assets. By tokenizing a good, sellers can trade with their surplus, and we can abstract the utility of a good. Buyers of a good can still exchange their goods for other goods of their choice, but also has the option of strategizing to maximize profit, and leverage -- similar to traditional financial markets.

\subsection{Market Making}
\hspace*{15pt}
Market Makers exist in every financial market. An individual, or entity, can provide market making services to the market. By setting up, and bootstrapping the markets continuation, an entity or group then become market makers. The Market Maker also has an inherent incentive to protect their value in the market. Market Makers must view the value given to a person seeking coins as proportional to the market value of the coins given to the person. 

\hspace*{15pt}
A useful analogy for this is a pawn shop exchanging dollars to a seller of a good being pawned. The pawn shop clerks inspect the good before exchanging value with the seller. This also incentivizes market marker competition to create and support a market network. 

\hspace*{15pt}
In the traditional financial markets, Market Makers are individuals, and entities who own enough capital to influence the markets. This creates a situation of exclusivity with respect to Market Making. Allowing more competition for fair market making widens the door for several other individuals and entities interested in support their own markets. 

\hspace*{15pt}
At the time of this paper, some of the largest cryptocurrency exchanges have began to offer Market Making services to their enterprise users. \(\texttt{AOS}\) enables Market Making as a Service (MMaaS) to bartering markets. This is one approach to tokenized bartering of physical goods. Capturing any metadata related to the good further enables the integrity of its representation within the network to better mirror reality, although there is no process today through which we can convert 100\% of a physical good into a digital good. 

\hspace*{15pt}
Another method could make use of a good holder demonstrating physical ownership of the good when they receive coins, and upon exporting their coins from the system, through clerks. A clerk is less incentivized to provide coins to an entity without being able to verify the entity possesses the value in physical goods. This approach allows clerks to be relieved from the duty of housing goods. More schemes for tokenizing assets exist through academic publications, and even patents \cite{a_patent}, but little is covered on the actual workflow of securely representing a general physical asset as a digital asset. However, there exists extensive academic literature on a posteriori processes once an asset becomes tokenized and digital.

\subsubsection{Representing physical goods}

\hspace*{15pt}
Unless the literature defines a universal way to automatically exchange an arbitrary unit of value with a physical good, a manual transaction must be accomplished. However, security vulnerabilities exists with every human transaction. 

\hspace*{15pt}
This manual approach to representing physical goods as digital goods is simply to gain access to a market's network through the exchange of a physical good. Simply participating in economic transactions only requires a trader possess an asset being used on that market's network. There is more work to be done in universal mechanisms to exchange value of a physical good for a digital good, but this approach is the most direct.

\hspace*{15pt}
If Market Maker \(A\) creates a market with rules too restrictive, or allocates themselves too many coins, potential market participants can choose Market Maker \(B\), or Market Maker \(n\), but for the same market. This creates an opportunity to create a network for anyone with access to commodity hardware. 

\subsection{Multichain Exchange}
\hspace*{15pt}
With many physical markets, there must exist many digital markets. A future with one global market currency seems \textit{unlikely}, but not \textit{impossible}. If multiple digital networks exist, enabling economic transactions to span across markets, and across networks is important. With \(\texttt{AOS}\), the goal is to create standalone programmatic markets, and cryptoeconomies. With the same underlying protocol fabric, we can connect multiple markets, despite them using different assets. Purposefully, we abstract away the \textit{Asset Exchange Mechanism}.

\hspace*{15pt}
If asset \(x\) from market \(X\) wants to be exchanged with asset \(y\) from market \(Y\), we must first create a transaction on each network to represent the asset being transferred, and another transaction on the other market signifying the paired-asset being received.

\hspace*{15pt}
A simple approach would be for the exchange to take into account the asset's current price, and the paired-asset's current price. Using these two values, we can provide a direct conversion. The conversion rate can be fixed, dynamic, or arrived upon through network consensus via participants from both networks -- this scheme is an arbitrary example, and can be built upon in future work.

\section{Implementation}
\hspace*{15pt}
To physically realize this system, we must translate philosophical details into actual implementation details. 

\subsection{Proposal Mechanism}
\hspace*{15pt}
The system makes use of the concept of a proposal, \(P\). In distributed computing, you can have a system in which a message is broadcasted with no intention of being responded to -- this can be seen as "fire and forget"; UDP is an example of a "fire and forget" protocol scheme. However, to achieve truly \textit{collaborative computing}, communication within a network should be \textit{communicative}. 

\hspace*{15pt}
Concretely, this means whenever a node transmits a message, a receiving node should respond with information, and the initial node should compute logic based on the receiving nodes response. For this purpose, any amendment to the system's "source of truth" is encapsulated within a proposal. 

\subsubsection{Defining Proposals}

\begin{figure*}[ht]
\centering
	\includegraphics[width=0.9\textwidth]{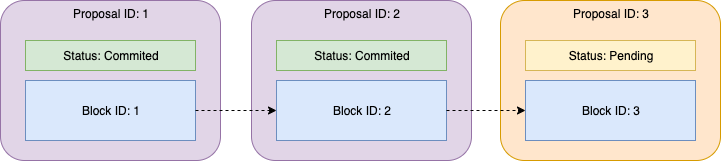}
	\caption{A proposal illustration, showing a chain of blocks being submitted by proposals}
	\label{proposals}
\end{figure*}

A furtherance of "time" in a decentralized system like Bitcoin is the creation of a new block. This adds another unit of \textit{time-sliced activity} onto the ever-growing ledger. Each transaction represents a change in the state of the system, which can also be viewed as a progression of time.

\hspace*{15pt}
In \(\texttt{AOS}\), a proposal is the trigger for each node to progress time, in the general sense. A node submits a proposal to the network, which must be verified by other nodes before being accepted by the network. Once all verification stages are complete, only then can the submitting node commit the proposal, which in term appends a new time-sliced block of activities to the global \textit{uniledger}.  

\subsection{Proposal Creation}
\hspace*{15pt}
This example of a simplistic mechanism for \textit{proposal creation} is implemented, and designed to be improved upon, and allow for custom agreement mechanisms to be created. 

\hspace*{15pt}
Let us say we have a network with 2 nodes participating in the proposal. To initiate the protocol, we simply invoke either node to produce the first proposal containing \(B_i = 0\), this is our \textit{genesis block}. To elect a node responsible for creating the next proposal, we define the function for \textit{proposal creator election}, \( \texttt{PCE} \) as

\begin{equation}
	\texttt{PCE}( \vert N \vert, B_i )
\end{equation}

thus, mapping a natural cartesian square, \(\field{N}^2\) to the node id set, \(N\),

\begin{equation}
	\texttt{PCE} :  \field{N} \times \field{N}  \rightarrow N_i
\end{equation}

\hspace*{15pt}
where \( \texttt{PCE} \) maps the cardinality of the set of network nodes participating in proposal creator election, a natural number \(\field{N} > 0\), and the current block height, \( B_i \), another natural number, \(\field{N} > 0\), to a node id, or \textit{node index}, also a natural number, \(\field{N} > 0\). \( N \) represents the set of nodes in the network of whom are being considered for creating the next proposal. From an individual node's perspective, this includes all of its peers, and itself. The scheme also assumes an ordered, and sequential set of node ids. To physically calculate \( \texttt{PCE} \), 

\begin{equation}
	\texttt{PCE} := ((B_i + 1) \: \texttt{mod} \: \vert N \vert) + 1
\end{equation}

\hspace*{15pt}
where we add \(1\) to \(B_i\) to represent the next block id, and we add \(1\) to the result assuming node ids, \(N_{id}\) begin at \(1\), not \(0\). The node then checks for whether or not its own node id, \(N_{id}\), is equal to the result of \texttt{PCE}. Effectively, this is checking whether or not \(N_{id}\) is congruent to the next \(B_i\) modulo the cardinality of nodes being considered or proposal creation, \(\vert N \vert\).  

\begin{equation}
	N_{id} = (B_i +1) \: (\texttt{mod} \: \vert N \vert)
\end{equation}

\hspace*{15pt}
The node elected to create the next proposal will always have a node id equal to the result of \(\texttt{PCE}\). This example is a round-robin approach to proposal creation among peers. Other methods for determining the creator of the next proposal can also be used instead of this basic example, both collaborative and not collaborative.

\subsection{Proposal State}
\hspace*{15pt}
The following states represent the stages of Practical Byzantine Fault Tolerance (\texttt{PBFT}) used for proposal agreement amongst participating nodes. This is a simple implementation of the \texttt{PBFT} protocol to reach agreement among peers.

\begin{enumerate}
	\item Created
	\item Response
	\item Resolution 
\end{enumerate}

\hspace*{15pt}
The protocol uses these three states to transition a submitted proposal through the network for agreement consideration. The physical implementation also has substates to assist with intermediary computations, but these substates fall into one of these main states, and may depend on the physical implementation details.  

\subsubsection{Created}
\hspace*{15pt}
The \textit{created} state coincides with \textit{pre-prepare} in \texttt{PBFT}, and signifies a node submitting a new proposal for consideration amongst participating nodes in the network.

\subsubsection{Response}
\hspace*{15pt}
The \textit{response} state coincides with \textit{prepare} in \texttt{PBFT}, and signifies a node informing the network of its acceptance or rejection of the submitted proposal. These response messages are transmitted by every node other than the node responsible for creating of the newly submitted proposal.

\subsubsection{Resolution}
\hspace*{15pt}
The \textit{resolution} state represents a node's commitment of the submitted proposal, as per the \textit{commit} state of \texttt{PBFT}. Every node participating in the agreement mechanism for a submitted proposal transmits a resolution signal to the rest of its peers.

\subsection{Proposal Agreement}
\hspace*{15pt}
The fundamental agreement mechanism is inspired by \texttt{PBFT} \cite{PBFTC}. This protocol is not dependent on any specific agreement mechanism, and can be implemented with an arbitrary agreement protocol, although network properties will differ with respect to the agreement protocol chosen.

\subsubsection{PBFT}
\begin{figure*}[ht]
\centering
	\includegraphics[width=0.7\textwidth]{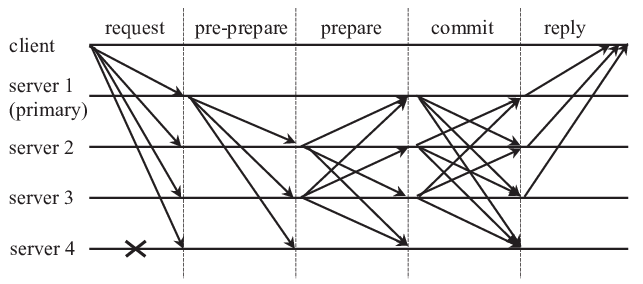}
	\caption{The protocol behavior for Practical Byzantine Fault tolerance (\texttt{PBFT})}
	\label{PBFT}
\end{figure*}

\hspace*{15pt}
The \texttt{PBFT} agreement protocol follows a "pre-prepare", "prepare", "commit" format. This is best illustrated by figure \ref{PBFT}.

\subsection{Proposal Verification}
\hspace*{15pt}
To verify a proposal, the process may be different depending on the use case. However, there are several common denominators regarding the verification of a submitted proposal. For example, a node may wish to check the hash between the submitted proposal, and the proposal existing before it was submitted, the previously submitted proposal. Another criteria may be to check if the submitted \(proposal_{id}\) is simply larger than the proposal last held by the node. Although naive, this example demonstrates a simple criteria that may or may not be important for a given use case.

Due to the differences in how proposals may be verified, and accepted, we demonstrate creating a generalized Proposal Verification Function \(\texttt{PVF}\). 

\begin{equation}
\texttt{PVF} : P \rightarrow \{\texttt{Accepted}, \texttt{Rejected}\}
\end{equation}

of which is physically implemented as,

\begin{equation}
\texttt{PVF} : P \rightarrow \{0, 1\}
\end{equation}

with \(0\), and \(1\) representing \(\texttt{invalid}\), and \(\texttt{valid}\) proposals respectively.

\subsubsection{Invalid Proposals}
\hspace*{15pt}
The cardinality of the set of reasons for which a submitted proposal is not accepted by peers is purposely minimal to mitigate complexity issues. In simplistic terms, a valid next proposal should have the correct next block index, and a correct proposal hash. A function to calculate the valid next block index , \(\texttt{VNBI}\) is defined as

\begin{equation}
\texttt{VNBI} : P \times \field{N} \rightarrow \{ 0, 1\}
\end{equation}

where the \(\field{N}\) is \(B_i\), and valid proposal hash as

\begin{equation}
\texttt{VPH} : P \rightarrow \{ 0, 1\}
\end{equation}

This results in \( \texttt{PVF} \) being computed as

\begin{equation}
\texttt{PVF} := \texttt{VNBI} \land \texttt{VPH}
\end{equation}

\hspace*{15pt}
This assumes the validity functions return 1 for valid proposals. If by physical implementation, \( \texttt{VNBI} \) and \( \texttt{VPH} \) represent functions that test for invalidity, meaning they return \(1\) for invalid proposals, then \( \texttt{PVF} \) is defined as 

\begin{equation}
\texttt{PVF} := \neg (\texttt{VNBI} \lor \texttt{VPH})
\end{equation}

These two definitions are essentially representing the same logical outcome of proposal validation. 

\subsection{Linking Proposals}
\hspace*{15pt}
Similar to how Bitcoin hashes a block header, containing the hash of its parent block header \cite{BTC}, we can link proposal history by hashing proposals. A proposal is inadvertently linked to a parent proposal because each block contains the hash of the proposal from which it was submitted. Thus, linking blocks, as covered in the next section, also inherently links proposals.

\section{Blocks in a chain}
\hspace*{15pt}
A shared ledger is the underlying source of truth in Bitcoin, Ethereum, and almost all (not all) modern blockchain-based protocols, sometimes referred to as \textit{satoshi-style} blockchains. This system uses the same concept for the same purpose, although we abstract away the logical details for generating blocks. This is because it is irrelevant to the conveyance of this system. An individual implementing group for this project may differ in how they produce blocks from another implementing group. The mechanism by which blocks are produced is not an \textit{implementation-breaking} details for \(\texttt{AOS}\). 

\subsection{Defining a Block}
\hspace*{15pt}
Time in the system is committed by creating \textit{time-sliced activity} blocks, similar to blocks in Ethereum \cite{ETH}, or Bitcoin \cite{BTC}. These blocks are cryptographically linked, and contain several important pieces of information. The included information can include a timestamp, the block hash, the node who committed the block, and the hash of the proposal responsible for triggering the block to be created, among other pieces of information.

\subsection{Creating Blocks}
\hspace*{15pt}
Depending on the decentralized protocol being examined, blocks are \textit{authored} differently within a protocol. In order to take advantage of the nuances within the decentralized protocol space, we wish to extend a generalized block generation function, \(\Phi\). Every proposal created contains a proposed block within its body. Regardless of implementation, the definition of the block production function is,

\begin{equation}
\Phi : P_{B_{\texttt{i}}} \rightarrow P_{B_{\texttt{i+1}}}
\end{equation}

Whereby \(P_{B_{\texttt{current}}}\) contains the block produced, \(\texttt{B}_{i+1}\). Here we index proposals by the block index in contains.

\subsection{Linking Blocks}
\hspace*{15pt}
In a similar fashion to Bitcoin, and \textit{hashed linked lists}, each block is cryptographically included in the hash of its child block. The purpose of hashing any data, \(D\), and including any other data, \(D^{\prime}\), while hashing \(D\), is to cryptographically prove \(D^{\prime}\) must have existed before \(D\) -- this is a property of composition. 

\begin{equation}
	\texttt{Hash}( \: \texttt{Hash}(\texttt{B}_{-1_{header}}) \:)
\end{equation}

Essentially, this is the same as the composite function,

\begin{equation}
(\texttt{H} \circ \texttt{H}) \: ( \texttt{B}_{-1_{header}} ) 
\end{equation}

and effectively proves that the inner hash function of \(  \texttt{B}_{-1_{header}} \) has to exist before the outer hash function is called (assuming \( \texttt{B}_{-1_{header}} \) is used in the hash). The amount of time that provably exists between the two functions being invoked is \textit{preserved}, assuming we can prove the amount of time existing between the inner and outer functions.

\section{Transactions}
\hspace*{15pt}
The limited capability of transactions from Bitcoin has led to modernization of decentralized protocols. This modernization has brought about the heavy usage and adoption of \textit{smart contracts}. However, contrary to popular belief, more advanced capabilities were always able to be computed by Bitcoin Script \cite{BTC}, provided a developer could write the desired operation into the simple scripting language, \textit{Bitcoin Script}. Nevertheless, Ethereum (and other protocols using smart contracts) have reintroduced to developers the concept that transactions for a protocol should be able to perform any \textit{turing complete} piece of functionality. 

\begin{figure}[ht]
\centering
	\includegraphics[width=0.2\textwidth]{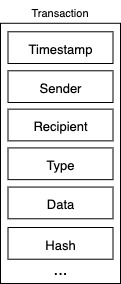}
	\caption{Components of a standard transaction}
	\label{tx}
\end{figure}

\hspace*{15pt}
It is due to this that \(\texttt{AOS}\) abstracts away the behavior of transactions. We generalize two types of transactions. One set of transactions, Type A, act as transactions submitted to the chain that \textit{await} invocation from other transactions, and another set of transactions, Type B, that are also submitted to the chain that \textit{invoke} Type A transactions. 

\subsection{Transaction Types}
\hspace*{15pt}
In Bitcoin, using Unspent Transaction Output (\(\texttt{UTXO}\)), \(\texttt{Type A}\) transactions are called \textit{outputs}, and \(\texttt{Type B}\) transactions are called \textit{Inputs} \cite{BTC}. With Ethereum, \(\texttt{Type A}\) are \textit{contract creation} transaction types, and \(\texttt{Type B}\) are \textit{Message Calls} \cite{ETH}. With Kunta, the protocol abstracts away transaction types, with \(\texttt{Type A}\) being \textit{female} transactions, and \(\texttt{Type B}\) being \textit{male} transactions \cite{KP}. Nevertheless, these two transaction types facilitate the nuances in behavior among all transactions handled by the system. Each transaction changes the state of the system, and each transaction type alters the state of the system in different ways.

\subsection{Signing}
\hspace*{15pt}
Every transaction submitted is signed by the sender using their private key, while their corresponding public key is public knowledge. The sender then encrypts the transaction using the public key belonging to the intended recipient. 

\hspace*{15pt}
Whenever the intended recipient of the transaction attempts to invoke the transaction directed to them, the recipient must first decrypt the transaction using their own private key, proving they own the key, of which is paired with the key used to encrypt the data on the first pass. This results in message signed by the sender. The intended recipient then verifies the transaction was sent by the actual sender by using the sender's public key for verification of the decrypted signature  -- this scheme ensures two things:

\begin{enumerate}
	\item The only peer allowed to invoke the transaction must own the private key corresponding to the public key used to encrypt the transaction 
	\item The peer who decrypts the transaction can verify the transaction was not sent by a peer who does not possess the private key used to initially sign the transaction
\end{enumerate}

\subsection{Transaction Algebra}
\hspace*{15pt}
Conventionally, developers design a \textit{virtual machine} with \textit{opcodes} dedicated to the execution upon submitted transactions. This section contains examples of using \textit{quasi}-boolean expressions to compute upon transactions, opposed to creating an arbitrary virtual machine for the task, as in Bitcoin \cite{BTC}, Ethereum \cite{ETH}, Kunta \cite{KP}, and other decentralized protocols. 

\subsubsection{Example 1}
\hspace*{15pt}
The most simple example of a Female transaction, \(\texttt{Tx}_f\), is a transaction containing an element from the binary set, \(\{0,1\}\). A corresponding Male transaction, \(\texttt{Tx}_m\), contains an element from the same set as well. Intuitively, this represents a situation where the sender of \(\texttt{Tx}_f\) is signaling their willingness to do something, or a "yes" represented as \(A\). This is a truth value of \(\texttt{True}\). The corresponding \(\texttt{Tx}_m\), also containing a 1, signals to the function their willingness, to do something, or a "yes".  The boolean function as a whole is

\begin{equation}
A \times B
\end{equation}

\hspace*{15pt}
Intuitively the function reads as "\(A\) and \(B\)", \(AB\), or \(A \land B\). The intuition of what the function result means to the network participants can be arbitrary. An example would be to transfer the ownership of an asset, controlled by the transaction, from \(A\) to \(B\). 

\begin{figure}[ht]
\centering
	\includegraphics[width=0.3\textwidth]{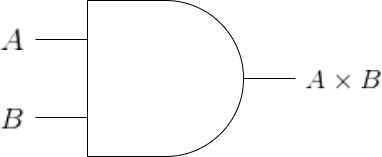}
	\caption{Transaction example 1, \(A \times B\),  transaction represented as a boolean circuit}
	\label{txexample1}
\end{figure}

\subsubsection{Example 2}
\hspace*{15pt}
This example uses a numerical value of units (which could represent a monetary value). This function would return the value to be subtracted from the senders account by using a conjunction on two variables, each submitted by a peer, and multiplying the truth value, \(\{0,1\}\), by the subtraction result. The amount to subtract is represented here by \(\Delta\). The following are equivalent:
\begin{equation}
A(B+C) \times \Delta
\end{equation}

\begin{equation}
\Delta(A(B+C))
\end{equation}

\begin{equation}
\Delta(AB+AC)
\end{equation}

\hspace*{15pt}
with \(\times\) representing multiplication, not the cartesian product, as both operands are not sets; \(A(B+C)\) is a truth value, and \(\Delta\) is a numerical value, a scalar. The \(\Delta\) would only be subtracted if \(A\) submits a confirmation, and either \(B\) or \(C\) submit confirmations. This results in \(\Delta\) being multiplied by 1, otherwise it is multiplied by zero, resulting in 0 units being added to the recipient's account, and zero being subtracted from the sender's account.

\begin{figure}[ht]
\centering
	\includegraphics[width=0.4\textwidth]{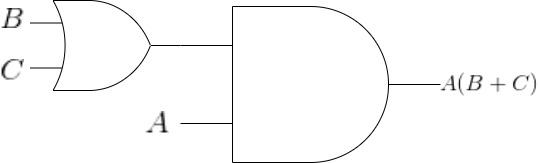}
	\caption{Transaction example 2, \(A(B+C)\), transaction represented as a boolean circuit}
	\label{txexample2}
\end{figure}

\subsubsection{Example 3}
To create an election, and compute votes, \(v\), submitted by a user, \(u\), for a ballot, \(b\), each vote is an element in \(V\).

\begin{equation}
u_{v_{b}} \in V_b \rightarrow \{u_{v_{b}}\} \cup V_b
\end{equation}

where, \(v \in \{0,1\} \). Votes from a user, for a ballot is

\[
    u_{v_{b}} = 
\begin{cases}
    1,		   & \text{if } u_{v_{b}} = 1\\
    0,              & \text{otherwise}
\end{cases}
\]

 An aggregate function to compute the votes can be,

\begin{equation}
f(b) = \sum_{u=1}^{\vert V_b \vert} u_{v_{b}}
\end{equation}

where \(\vert V_b \vert \) is the total votes for \(b\), \(u_{v_{b}}\) is the value corresponding to a vote from \(u\) for \(b\). Finally, we index \(v\), by \(b\), \(v_b\),  for a ballot, where \(b \in B\), with \(B\) being the set of all candidates. To complete an election, if all candidates must confirm election results, the winner, \(w\), is defined by

\begin{equation}
w = \operatorname*{argmax}_{b} f(b)
\end{equation}

and the transaction logic simply is

\begin{equation}
w(ABC ... n)
\end{equation}

where \(A\), \(B\), \(C\), etc are the confirmations of individual candidates/ballots. The winner is then \(B_{w(ABC ... n)}\).

\begin{figure}[ht]
\centering
	\includegraphics[width=0.25\textwidth]{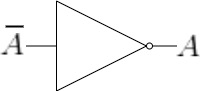}
	\caption{Transaction example 3, \(A\),  transaction represented as a boolean circuit}
	\label{not}
\end{figure}

\hspace*{15pt}
Implementation would be a transaction containing a placeholder for a single value. The single value would be a bit, representing a vote, \(\{0,1\}\), initially set to 0. If we represent it as a circuit, we could use an inverter on the negation of the bit, shown in figure \ref{not}, which is logically equivalent to just taking the value of the bit. For illustrative purposes, \(A\) just represents a value stored in a transaction.

\subsubsection{Conjunction}
\hspace*{15pt}
We perform a conjunction between transaction examples 1, and 2 above, (\(t_1\),\(t_2\) respectively), by,

\begin{equation}
R_{ABC} \equiv A_{t_1}B_{t_1} \land A_{t_2}(B_{t_2}+C_{t_2})
\end{equation}

this truth value is \(\texttt{True}\) only if both transactions are true.

\begin{equation}
R_{ABC} \times \Delta
\end{equation}

\hspace*{15pt}
The output of the boolean statements are used as variables in the algebraic statement. In this example illustrated, the return value of the entire transaction conjunction would be the amount subtracted from the sender's account, and added to the recipient's account. Using these two layers of algebra, we can compute logic for single transactions, and multiple transactions.

\subsection{Designing Mechanisms}
\hspace*{15pt}
To keep it simple, we focus on two-party mechanisms. We say Agent \(i\), \(A_i\), represents one of two diametrically opposed parties, and acts as a lobbyist. The agent will experience \textit{political pressure} from the parties they represent. The \textit{political pressure functions} are the diametrically opposed functions.

\begin{figure}[ht]
\centering
	\includegraphics[width=0.3\textwidth]{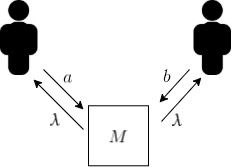}
	\caption{The interaction between two agents, \(A\), through mechanism \(M\)}
	\label{mechanisms_simple}
\end{figure}

\begin{figure*}[ht]
\centering
	\includegraphics[width=0.8\textwidth]{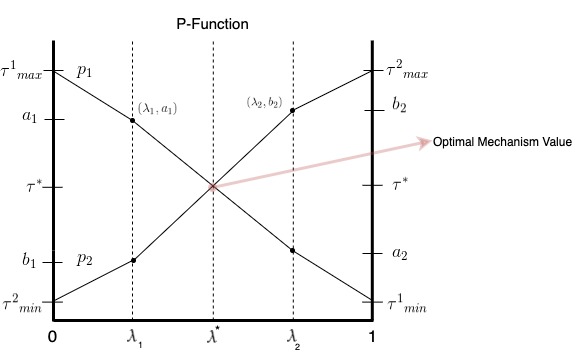}
	\caption{Abstract representation of a mechanism, \(M\), between two diametrically opposed agents, and their environment. \(\lambda\) is the value controlled by the mechanism, where \(a\), and \(b\) are the private messages from the agents to \(M\)}
	\label{mechanisms_general}
\end{figure*}

\hspace*{15pt}
Illustrated in figure \ref{mechanisms_general}, we define a general mechanism with two diametrically opposed agent p-functions. We refer to the functions \(P_i\), \textit{political action} functions, or \textit{p-functions}. For simplicity, we treat the p-functions as primitives; we make two assumptions about them. First, we assume that the function \(P_1\) takes values in the interval \cite{DEM}

\begin{equation}
[\tau^i_{min}, \tau^i_{max}]
\end{equation}

where i = 1,2, since there are two agents. The value \( \tau \) is the output of the environment, from \( M \), or the output of the mechanism, given the responses from the agents, \( A \). A p-function, \(P_1\), is uniquely specified by two numbers, \( a_1 \), and \( a_2 \). Similarly, \( P_2 \) is characterized by two numbers, \( b_1 \), and \( b_2 \). Thus, an environment consists of a possible pair of functions \( (P_1, P_2) \). It is specified by four numbers, \( \theta = (a_1, a_2, b_1, b_2) \). The set, \( \Theta = \Theta^1 \times \Theta^2 \) of environments is the set of all \( \theta = (\theta^1, \theta^2) \) that satisfy these conditions

\begin{equation}
\tau^1_{max} > a_1 > a_2 > \tau^1_{min}
\end{equation}

and

\begin{equation}
\tau^2_{min} < b_1 < b_2 < \tau^2_{max}
\end{equation}

Thus, 
\begin{equation}
\Theta^1 = \{  (a_1, a_2) : \tau^1_{max} > a_1 > a_2 > \tau^1_{min} \}
\end{equation}

and 
\begin{equation}
\Theta^2 = \{  (b_1, b_2) : \tau^2_{min} < b_1 < b_2 < \tau^2_{max} \}
\end{equation}

\hspace*{15pt}
The job of \(M\) is to accept input from both agents, \(a\), and \(b\). Value \(a\) is a set containing elements in the codomain of the p-function for \(A_1\). In the simple sense,

\begin{equation}
a \equiv \{a_1, a_2\}
\end{equation}

and \(a\) is the response of \(A_1\) to \(M\) returning \(\lambda\). Note that each \textit{p-function}, is either monotonically increasing, or monotonically decreasing. This is convenient in order for the p-functions to be diametrically opposed.

\subsubsection{\(M\) for Pricing}
\begin{figure}[ht]
\centering
	\includegraphics[width=0.5\textwidth]{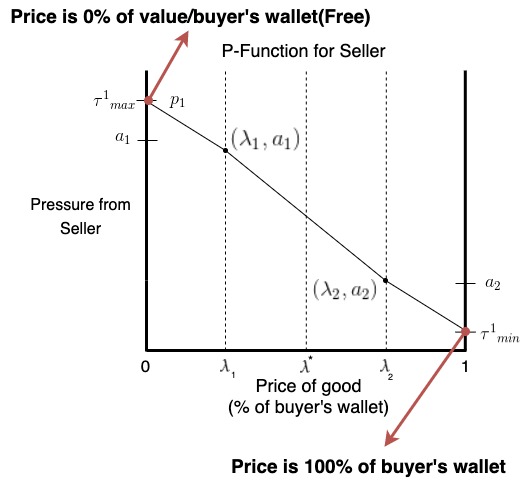}
	\caption{Mechanism for seller, \(M_{seller}\), where \(\lambda\) is the price of the good being exchanged. At \(\lambda\) = 1, the price becomes 100\% of the buyer's reserves, or the seller's/market's perceived value of the good, \(g\)}
	\label{mechanisms_seller}
\end{figure}

\hspace*{15pt}
For example, we say the mechanism, \(M\) regulates the price of good, \(g\), for an exchange of \(g\) from agent 1 to agent 2.    For a buyer and a seller, the price of \(g\) can be set by buyers. However the price can also be set by a mechanism. The buyer and seller serve as the two diametrically opposed parties, since the buyer wants to minimize price, and the seller wants to maximize it. 

\hspace*{15pt}
The diametrically opposed p-functions represent the political pressures exerted upon the agents, by the buyer and seller of \(g\). If \(\lambda\), the price of \(g\) in the transaction, is 0, the political pressure felt by the seller will be high, but the buyer will be elated at \(g\) being \textit{free}, feeling little or no pressure. Once a value of \(\lambda\) is proposed by \(M\), \(a\) and \(b\) both report their p-function values at \(\lambda\). In this example, the goal function is to balance, or optimize \(\lambda\). This is just an example goal function over both p-functions, and much more advanced goal functions can be constructed using the same principles.

\hspace*{15pt}
Figure \ref{mechanisms_seller} shows the \textit{p-function} for the seller of \(g\). When \(\lambda\) takes the value of 0, the political pressure felt by the agent representing the seller is maximal. Intuitively, this is because the seller of \(g\) is not economically incentivized to give \(g\) away for free. When \(\lambda\) takes the value of 1, this represents the price of \(g\) being equivalent to the buyer's total reserves. The opposition to this p-function is the p-function of agent 2, \(a_2\), of which is representing the buyer. If \(\lambda\) is 1, then the political pressure imposed by the buyer onto \(a_2\) is maximal. \(\tau^2_{max}\) can be viewed as 100\% of the value of \(g\), or 100\% of the buyer's reserves.

\subsection{Use in privacy}
\hspace*{15pt}
With several approaches towards handling \textit{privacy}, of which do not have to be mutually exclusive. However, we examine them both independently.

\subsubsection{Confidential Transactions}
\hspace*{15pt}
Hiding information in a transaction would occur by masking true information with blinding factors from the algebraic statements. By doing this, onlookers cannot feasibly know the variables used in the computation without the involved parties revealing the information. For example, \(A\) could represent something sensitive for a node, but since the value is in \(\mathbb{R}\), it can be hidden by several combinations of masking, and abstraction.

\hspace*{15pt}
If we have boolean statements, the size of the search space for a nefarious peer would be proportional to the amount of variables in the statements. Each variable can only be from the set \(\{0,1\}\). With 2 variables, the peer would have to guess \(2^2\) combinations. 

\hspace*{15pt}
A security margin is a measure of the required effort needed to break a systems security measure(s) -- estimating the hardness of breaking the system. If you have a high enough security margin, the nefarious actor may not have the opportunity to perform the necessary amount of attempts to correctly guess the blinding factors, as they may be penalized, punished, or blocked after \(0 < a < 4\) attempts. 

\hspace*{15pt}
One method to circumvent this is to add variables to the statements, while the sender, and receiver know the correct negations to be made. This yields \(2^n\) combinations for the bad actor to guess. This can be used for components of the computation executed upon during a state change caused by a transaction being invoked. Following the scheme set forth by Pedersen, \cite{PED}, the commiter commits themself to an \( s \in \mathbb{Z}_q \) by choosing \(t \in \mathbb{Z}_q\) at random and computing

\begin{equation}
E(s,t) = g^s h^t
\end{equation}

\hspace*{15pt}
Such a commitment can later be opened by revealing \(s\) and \(t\). It is very easy to prove and show that \(E(s, t)\) reveals no information about s, and that the committer cannot open a commitment to \(s\) as \(s' \neq s\) unless he can find \(log_g(h)\). For any \( s \in \mathbb{Z}_q \), and for randomly uniformly chosen \( t \in \mathbb{Z}_q \), \(E(s, t)\) is uniformly distributed in the group \(G_q\). Once the commiter makes \(E(s, t)\) public, and reveals \(s\) and \(t\), the verifier is given \(E(s, t)\), \(s\), and \(t\). The verifier can then check the commitment for integrity. The integrity of \(E(s, t)\) holds as long as the commiter adheres to the scheme. This scheme can be used to add a layer of privacy to a transaction.

\subsubsection{Differential Privacy}
\hspace*{15pt}
Modeled after randomized processes known as \textit{randomized response}, Differential privacy \cite{DP} is useful for computing approximations over populations. For example, to compute an approximation of whether or not a population of people in a country prefer option A, the higher our sample size, the more we can tolerate untruthful answers from respondents. 

\hspace*{15pt}
Regarding decentralized protocols, we can only tolerate approximations for a subset of problems. There exists another subset of problems for which we must compute deterministic truths. RSMs typically execute deterministic virtual machine (VM) computations upon parameters, values, and operations.

\hspace*{15pt}
By using differential privacy, privacy comes from the plausible deniability of any outcome \cite{DP}. Accuracy comes from an understanding of the noise generation procedure. Non-triviality says that there exists a query and two databases that yield different outputs under this query. Changing one row at a time we see there exists a pair of databases differing only in the value of a single row, on which the same query yields different outputs. An adversary knowing that the database is one of these two almost identical databases learns the value of the data in the unknown row. Also, differential privacy does not guarantee that what one believes to be one's secrets will remain secret.

\hspace*{15pt}
A mechanism for differentially private transactions can be viewed from various perspectives. For example, If we wish to provide  \(\epsilon\)-differential privacy for a transaction, we would have to have a set of transactions that may be invalid, or generated by the privacy mechanism. This would enable participants with plausible deniability regarding whether or not a transaction belonging to them has been generated by the privacy mechanism, or themselves directly. 

\hspace*{15pt}
Say \(\epsilon\) be a positive real number, $\mathbb{R}$, and \texttt{A} be an algorithm, chosen at random, that expects a dataset as input. Let \(\texttt{im}\, {\mathcal {A}}\) be the image of \texttt{A}. The algorithm \(\mathcal {A}\) provides \(\epsilon\)-differential privacy if, for all datasets, \(D_{1}\) and \(D_{2}\), differing only on on a single record, and all subsets \(S\) of \(\texttt{im} \, {\mathcal {A}}\):

\begin{equation}
 \Pr[{\mathcal {A}}(D_{1})\in S]\leq \exp \left(\epsilon \right)\cdot \Pr[{\mathcal {A}}(D_{2})\in S],
\end{equation}

\hspace*{15pt}
If a participant's transaction is in either database, \(D\), the participants can claim the transaction is not truthful. This is only one approach to applying differential privacy to transactions directly. Other schemes may include applying \(\epsilon\)-differential privacy to the computations themselves, and not the transaction as a whole.

\section{Further work}
\hspace*{15pt}
Moving forward, this protocol supports custom block production functions, agreement mechanisms, and transaction invocation functions to be implemented. The goal of this work is to satisfy the minimum requirements suitable to implement a distributed protocol for proposal agreement. With this system, groups can create their own markets, with their own rules, using a system built with a minimum amount of constraints necessary to generate distributed economic collaboration, and commerce. This system is intentionally developed to be built upon, by us, and other research teams.

\subsection{Research Horizon}
\hspace*{15pt}
Since scientific work is ever-expanding, the items on our horizon fall into two categories. \textit{privacy}, and \textit{customization}. These categories are independent of the standard enhancements forthcoming for software-based development projects.

\subsubsection{Privacy}
\hspace*{15pt}
Several items on our research horizon pertain to guarantees of privacy. Mainly, using \textit{arguments from ignorance} to enhance privacy capabilities of a system. Privacy is a big concern, and we wish to enhance this work by experimenting with the tradeoffs between enhanced privacy, the message-space size, compute time, and storage space required (\textit{disk} and \textit{memory}). There are several research teams working on this, both in pure theory, and in practice. Our implementation of confidential transactions is but only one example of using arguments from knowledge.

\subsubsection{Universally Modular}
\hspace*{15pt}
Another goal of this work is to mitigate conflicts of implementation. There exists a subset of implementers who wish to control parameter(s) of the RSM. Maximizing the parameters allowed to be extended, and altered, without sacrificing underlying security features is a primary goal. There also exists an opportunity to extend development powers to developers for the underlying security features themselves. It is common for a protocol in this space to guard their underlying protocol features, opposed to embracing customization. In this spirit, we build from methodologies demonstrated in the Kunta paper \cite{KP}. 

\subsubsection{Developer Friendliness}
\hspace*{15pt}
The definition of \textit{Developer Friendliness} should be widely reexamined. Simply because a project is \textit{open source}, does not automatically translate into the same project being \textit{developer friendly}. Developer friendliness is a measure of how much the project has taken into account the average developer using the tools used in the project. 

\hspace*{15pt}
This includes the programming language, and the efficiency of the average developer using it. It is good if a project is open source, but it is \textit{suboptimal} if the esoteric nature of the code base confuses developers wishing to customize its core features. More work is needed to be done if the industry is to make the code bases of decentralized protocols more developer friendly.

\end{document}